\documentclass[floats,floatfix,showpacs,amssymb,prl,twocolumn,superscriptaddress,nofootinbib]{revtex4-1}

\usepackage{amssymb,amsmath,verbatim,mathtools,needspace,enumitem,etoolbox,graphicx,physics,microtype,afterpage,bm}

\usepackage[dvipsnames, usenames]{xcolor}

\definecolor{linkcolor}{rgb}{0.0,0.3,0.5}
\usepackage[unicode, colorlinks=true, linkcolor=linkcolor, citecolor=linkcolor, filecolor=linkcolor,urlcolor=linkcolor, pdfusetitle]{hyperref}
\usepackage[all]{hypcap}
\usepackage[T1]{fontenc}
\usepackage[utf8]{inputenc}
\usepackage{tabularx}
\usepackage{float}
\usepackage{url}
\allowdisplaybreaks
\graphicspath{{./figure/}}

\def\aj{\rm{AJ}}
\def\araa{\rm{ARA\&A}}
\def\apj{\rm{ApJ}}
\def\apjl{\rm{ApJ}}
\def\apjs{\rm{ApJS}}
\def\aap{\rm{A\&A}}
\def\aapr{\rm{A\&A~Rev.}}

\def\mnras{\rm{MNRAS}}
\def\prd{\rm{PRD}}
\def\prl{\rm{PRL}}

\def\pasp{\rm{PASP}}

\def\ssr{\rm{Space~Sci.~Rev.}}

\newcommand{\jhu}{\affiliation{Department of Physics and Astronomy, Johns Hopkins University, 3400 N. Charles
Street, Baltimore, MD 21218, USA}}
\newcommand{\flatiron}{\affiliation{Center for Computational Astrophysics, Flatiron Institute, New York, NY 10010, USA}}
\newcommand{\TAPIR}{\affiliation{TAPIR, California Institute of Technology, Pasadena, CA 91125, USA}}
\newcommand{\Carnegie}{\affiliation{The Observatories of the Carnegie Institution for Science, Pasadena, CA 91101, USA}}

\definecolor{rb4}{HTML}{27408B}

\begin{document}

\title{Joint constraints on the field-cluster mixing fraction, common envelope efficiency, and globular cluster radii from a population of binary hole mergers via deep learning}


\author{Kaze W. K. Wong} 
\email{kazewong@jhu.edu}
\jhu

\author{Katelyn Breivik}
\flatiron

\author{Kyle Kremer}
\TAPIR
\Carnegie

\author{Thomas Callister}
\flatiron

\date{\today}

\begin{abstract}
The recent release of the second Gravitational-Wave Transient Catalog (GWTC-2) has increased significantly the number of known GW events, enabling unprecedented constraints on formation models of compact binaries.
One pressing question is to understand the fraction of binaries originating from different formation channels, such as isolated field formation versus dynamical formation in dense stellar clusters.
In this paper, we combine the \texttt{COSMIC} binary population synthesis suite and the \texttt{CMC} code for globular cluster evolution to create a mixture model for black hole binary formation under both formation scenarios.
For the first time, these code bodies are combined self-consistently, with \texttt{CMC} itself employing \texttt{COSMIC} to track stellar evolution.
We then use a deep-learning enhanced hierarchical Bayesian analysis to constrain the mixture fraction $f$ between formation models, while simultaneously constraining the common envelope efficiency $\alpha$ assumed in \texttt{COSMIC} and the initial cluster virial radius $r_v$ assumed in \texttt{CMC}. 
Under specific assumptions about other uncertain aspects of isolated binary and globular cluster evolution, we report the median and $90\%$ confidence interval of three physical parameters $(f,\alpha,r_v)=(0.20^{+0.32}_{-0.18},2.26^{+2.65}_{-1.84},2.71^{+0.83}_{-1.17})$. 
This simultaneous constraint agrees with observed properties of globular clusters in the Milky Way and is an important first step in the pathway toward learning astrophysics of compact binary formation from GW observations.
\end{abstract}

\maketitle

\textbf{\textit{Introduction}} The number of gravitational-wave (GW) detections is growing at an accelerating pace since the first detection~\cite{Abbott:2016blz}.
The LIGO-Virgo-KAGRA Collaboration (LVKC) recently released the second Gravitational-Wave Transient Catalog (GWTC-2), which includes $39$ events from the first half of of the third observational run (O3)~\cite{Abbott:2020niy}.
The number of GW events in this new catalogue is $\sim 4$ times the number of events from the first two observational runs combined~\cite{LIGOScientific:2018mvr}, allowing for 
more sensitive explorations of their collective mass, spin and merger redshift distributions~\cite{Abbott:2020gyp}.
As the statistical uncertainties in these distributions continue to drop with the growing number of detections, the population of GW events provides a unique, and increasingly powerful, avenue to probe a wide range of topics, including fundamental physics~\cite{Abbott:2020jks}, cosmology~\cite{Abbott:2019yzh,Palmese:2020aof}, and astrophysics~\cite{Farmer:2020xne,Bouffanais:2020qds,Callister:2020arv}.

One of the most pressing questions in GW population analyses is which binary formation channels generate the observed GW events.
In particular, it is expected that compact binary mergers may arise via isolated binary evolution in the stellar field or dynamical assembly in dense clusters, although a variety of other channels have also been theorized~\citep[e.g.][]{bird_did_2016,antonini_precessional_2018,mckernan_constraining_2018}. 
In the isolated field scenario, compact binary mergers are the end result of stellar binary evolution~\cite{Dominik:2012kk,Belczynski:2017gds}.
On the other hand, events from dynamical formation scenarios are formed through multi-body encounters in dense environments, such as globular clusters, young stellar clusters, or galactic nuclear clusters~\cite{Rodriguez:2017pec,DiCarlo:2019pmf}.
While the precise distributions of compact binaries originating from each scenario are not known, field and cluster channels generally differ in their predictions regarding spin, eccentricity, and component mass distributions of binary mergers.
Due to mass transfer and tidal alignment, binaries resulting from isolated formation tend to have component spins aligned with their orbital angular momenta,
whereas dynamically formed binaries are believed to have an isotropic distribution of spins~\cite{Rodriguez:2016vmx,Farr:2017uvj,Qin:2018vaa,Bavera2020}.
Hierarchical mergers are likely only possible in a dynamical-formation scenario~\cite{Gerosa:2017kvu,Doctor:2019ruh,Rodriguez:2019huv,McKernan2020},
which could result in black holes with masses in the theorized pair-instability supernova (PISN) gap (commonly known as the ``upper mass gap")~\cite{Woosley:2016hmi,Farmer:2019jed}.

Some studies have already sought to use the observed masses and spins of compact binary mergers to infer the mixing fraction between the two formation channels~\cite{Bouffanais:2019nrw,Safarzadeh:2020jsc,Abbott:2020gyp}.
Such studies, however, generally adopt heuristic models for expected spin distributions, or ignore possible variation within individual channels that may be correlated with the inferred mixing fractions.
Moreover, although the mixing fraction itself is an important question, understanding the mixing fraction alone yields little insight on the physics of each underlying formation channel.
Thus, it is important to infer the mixing fraction and channel-specific parameters \emph{jointly}.

In this \textit{Letter}, we create a mixture model of merging binary black holes (BBHs) from isolated binary evolution using the binary population synthesis code \texttt{COSMIC} and from globular clusters (GCs) using the GC evolution code \texttt{CMC}. We infer the properties of this mixture model by applying a deep learning enhanced hierarchical Bayesian modeling framework on the GWTC-2 BBH catalogue. A review of the data analysis method is given in the Supplemental Material.



\textbf{\textit{Isolated binary evolution with}} \texttt{COSMIC}. We generate a cosmological population of BBH mergers originating from isolated binary evolution using \texttt{COSMIC}\cite{Breivik2020}, which is based on an updated version of \texttt{BSE} \cite{Hurley2002}. See \citep{Breivik2020} for a comprehensive summary of all upgrades currently employed in \texttt{COSMIC} and \cite{Zevin2020} for a discussion of the prescriptions which most heavily impact compact-object formation. We assume that massive stars are initially distributed with masses following a power law with index $\alpha = -2.3$ \cite{Kroupa01} and $70\%$ of them have companions with mass ratios distributed uniformly \cite{Sana12}. We distribute initial orbital periods with a power law in $\log\,P_{\rm{orb}}$ with index $\pi=-0.55$ and $\log\,P_{\rm{orb}}\in[0.15,5]$ and initial eccentricities with a power law with index $\xi=-0.45$. 

One of the largest uncertainties in binary evolution is the amount a binary's orbital separation shrinks as aresult of common envelope evolution \cite{Ivanova2013}. \texttt{COSMIC} employs the $\alpha\lambda$ prescription to parameterize how efficiently orbital energy is used in unbinding the stellar envelope, where $\lambda$ is the envelope binding energy and is calculated following Appendix A of \cite{Claeys2014} assuming no contributions from ionizing energy. Previous studies suggest a wide range of ejection efficiencies varying from $\alpha=0.25-5$ for a wide variety of stellar masses \citep[e.g.][]{Zorotovic2010, DeMarco2011, Fragos2019}. To capture this uncertainty, we ran $8$ separate models ($\alpha=0.25,0.5,0.75,1,2,3,4,5$) each with $16$ metallicity bins spaced logarithmically between $Z_{\odot}/200$ and $2Z_{\odot}$, where $Z_{\odot}=0.017$ \cite{Grevesse1998}. The binary evolution model, except for the variation of envelope ejection efficiency, is identical to that of \cite{Zevin2020} which consistently produces local ($z<0.01$) merger rates consistent within a factor of 2 to the $90\%$ credible interval of the observed rates from the GWTC-1 catalog for $\alpha=1,5$.

To generate a cosmological population of merging BBHs, we use the redshift-dependent star formation history and metallicity evolution of \cite{Madau2017} and assume \emph{Planck 2015} cosmological parameters: $H_0 = 68$ km s$^{-1}$ Mpc$^{-1}$, $\Omega_{m} = 0.31$ and $\Omega_{\Lambda} = 0.69$~\cite{2016A&A...594A..13P} as implemented in \texttt{astropy} \citep{astropy:2013, astropy:2018}. Similar to \cite{Zevin2020}, we assume a truncated log-normal distribution of metallicities with $\sigma=0.5$\,dex following \cite{Bavera2020}. We break the star formation into $100$ linearly spaced redshift bins with $z\in[0,15]$ and calculate the number of BBHs formed, weighted by their metallicity, by normalizing the total mass of stars from our simulated population to the total amount of star formation in each redshift bin. We then record the lookback time and redshift of each BBH merger to create a catalog of all merging BBHs for redshifts $z<15$. 

\textbf{\textit{Globular cluster evolution with}} \texttt{CMC}. We use $N$-body simulations presented in the \texttt{CMC Cluster Catalog} \citep{Kremer2020} to simulate GC evolution. These simulations were computed using \texttt{CMC} (for \texttt{Cluster Monte Carlo}) \citep{Joshi2000,Pattabiraman2013}, a H\'{e}non-type Monte Carlo code which includes various processes relevant to BH binary formation including two-body relaxation \citep{Joshi2000}, three-body binary formation \citep{Morscher2015}, direct integration of small-$N$ resonate encounters \citep{Fregeau2007,Rodriguez2018}, and stellar/binary evolution. For the latter, \texttt{CMC} uses updated versions of \texttt{SSE} and \texttt{BSE} \citep{Hurley2000,Hurley2002}, identical to those used in \texttt{COSMIC}, only varying in choices of binary evolution prescriptions.

In the \texttt{CMC Cluster Catalog}, four key cluster parameters are varied between the different simulations: initial number of stars per cluster ($N/10^5 = 2, 4, 8, 16$), initial virial radius ($r_v/\rm{pc} = 0.5, 1, 2, 4$), metallicity ($Z/Z_{\odot} = 0.01,0.1,1$), and radial position within a (Milky-Way-like) galactic potential ($R_{\rm{gc}}/\rm{kpc}$=2,8,20). Collectively, this simulation suite covers the full parameter space of the Milky Way GCs and captures the formation of a variety of astrophysical objects including GW sources, X-ray binaries, millisecond pulsars, and blue stragglers.

By coupling a cluster age distribution model from \citep{ElBadry2019} with the BBH merger delay time distributions gathered from the \texttt{CMC} models, a realistic distribution of dynamical BBH merger times can be assembled. In \citep{Kremer2020}, this method was used to estimate a BBH merger rate of roughly $20\,\rm{Gpc}^{-3}\,\rm{yr}^{-1}$ in the local universe (assuming all $N,\,r_v,\,Z,\,$and $R_{\rm{gc}}$ values are equally weighted), consistent with similar rate estimates from other recent studies \citep[e.g.,][]{RodriguezLoeb2018,FragioneKocsis2018,AntoniniGieles2020}.

While \texttt{COSMIC} can produce nearly arbitrarily large catalogs of compact binary mergers, the output of \texttt{CMC} is limited by the relatively high computational costs associated with dynamical $N$-body simulations \citep[e.g.,][]{Pattabiraman2013}. As a result, the total number of GW events from our set of \texttt{CMC} models is of order $10^4$ compared to $\sim10^6$ from each \texttt{COSMIC} model.
To mitigate the presence of small-scale fluctuations due to finite sampling, which may inadvertently be learned and reproduced by our emulator, we smooth the binary mass distributions given by each \texttt{CMC} simulation with a gaussian kernel using \texttt{scipy.stats.gaussian\_kde}.
The smoothing preserves physical features in the mass functions, such as high-mass ``bumps'' due to repeated mergers, while smoothing out small unwanted scale fluctuations.

\textbf{\textit{Mixture model}}. In an ideal scenario, a single model would be used to simultaneously predict both the observable properties and the merger rates of compact binaries across all different formation channels.
In this case, the model would directly predict the mixing fraction between channels with no additional free parameters.
In practice, the compact binary merger rates given by both \texttt{COSMIC} and \texttt{CMC} are highly uncertain and possibly subject to severe systematic bias.
Therefore
we deliberately introduce a free parameter $f$ controlling the mixing fraction:
\begin{align}
p(m_1,m_2,z|\alpha,r_v,f) &= f p_\text{\tiny COSMIC}(m_1,m_2,z|\alpha)\nonumber\\
&+(1-f) p_\text{\tiny CMC}(m_1,m_2,z|r_v),
\label{eq:mixingFraction}
\end{align}
where $p_\text {\tiny COSMIC}(m_1,m_2,z)$ and $p_\text {\tiny CMC}(m_1,m_2,z)$ are the probability densities on the primary mass $m_1$, secondary mass $m_2\leq m_1$, and redshift $z$ of BBH mergers predicted by \texttt{COSMIC} and \texttt{CMC}, respectively.
This choice 
ensures that our conclusions are physically informed by the \textit{shapes} of the observed mass and redshift distributions, and not on the \textit{rates} of 
mergers.
The parameters of the mixture model are summarized in table \ref{Tb:parameters}.

\begin{table}[h]
\caption{Event parameters and hyper-parameters used in this work.}
\begin{tabularx}{\columnwidth}{l X}
\hline
\hline
Event parameters $\bm{\theta}$ &  \\
\hline
$m_1 \in [2.5,100]\ M_{\odot}$ & Source-frame primary mass of the binary \\
$m_2 \in [2.5,100]\ M_{\odot}$ & Source-frame secondary mass of the binary \\
$z \in [0,1.2]$ & Redshift of the binary\\
\hline
Hyper-parameters $\bm{\lambda}$ &  \\
\hline
$\alpha \in [0.25,5]$ & Common envelope efficiency \\ 
$r_v \in [0.5,4]\ {\rm pc}$ & Initial cluster virial radius \\
$f  \in [0,1] $ & Fraction of binaries from field-formation channel \\
\hline
\hline
\end{tabularx}
\label{Tb:parameters}
\end{table}

\begin{figure}
\includegraphics[width=\columnwidth]{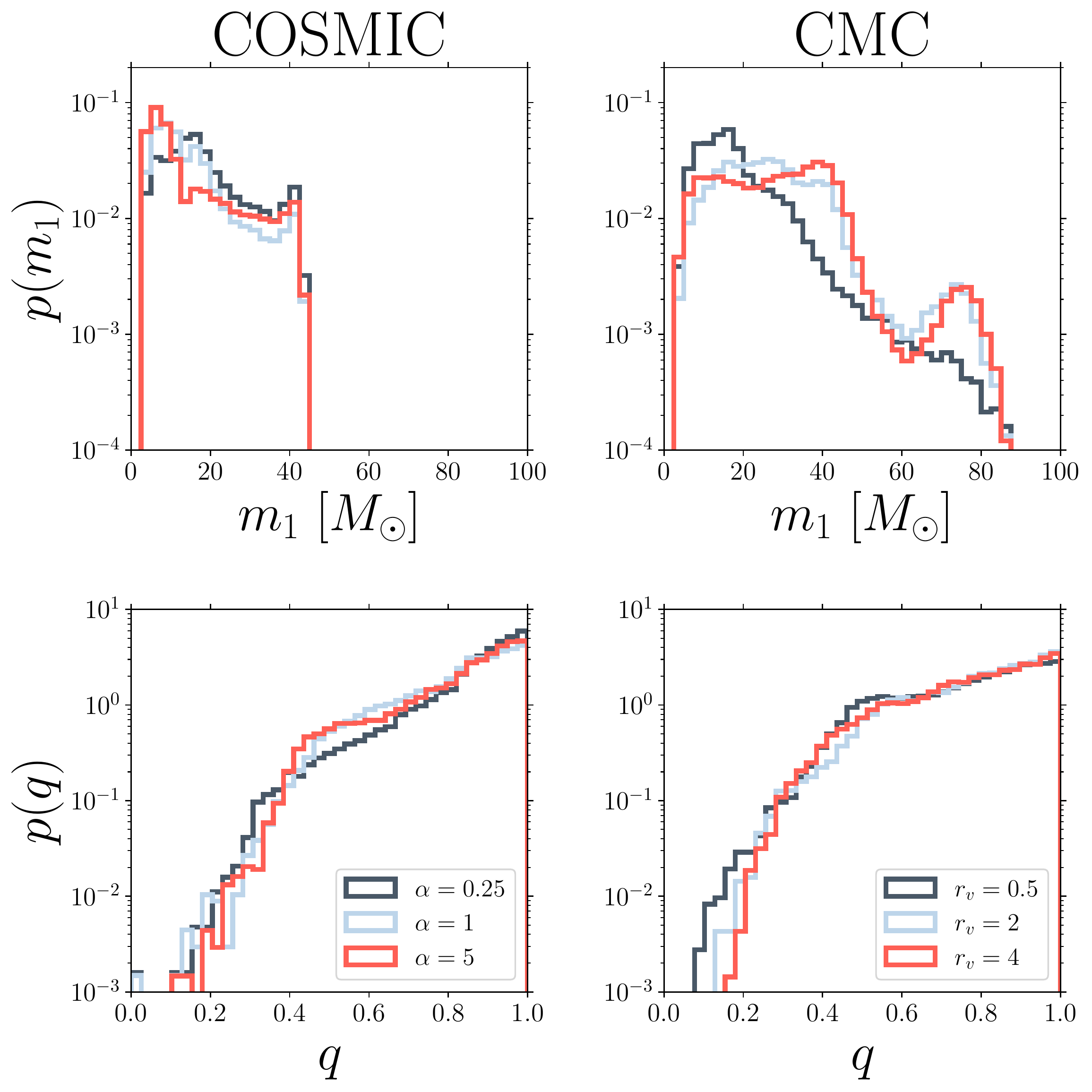}
\caption{Distributions of source-frame primary mass and mass ratio for all merging BBHs from the \texttt{COSMIC} and \texttt{CMC} models.
The left column shows the distributions for different common envelope efficiencies and merging BBHs from \texttt{COSMIC} only,
and the left column shows the distributions for different initial virial radii when we only consider BBH mergers from \texttt{CMC}.
}
\label{fig:m1q_Distribution_plots}
\end{figure}

Each combination of hyper-parameters 
yields a different distribution of primary masses and mass ratios $(q=m_2/m_1)$.
Several different examples are shown in Fig.~\ref{fig:m1q_Distribution_plots}, varying the common envelope efficiency $\alpha$ in \texttt{COSMIC} (left-hand side) and the initial cluster virial radius $r_v$ in \texttt{CMC} (right-hand side).
Larger common envelope efficiencies
produce relatively lower primary masses, while clusters with larger $r_v$ retain more massive BHs at late times and thus exhibit a high-mass peak around $80\,M_\odot$ due to repeated BH mergers.

\begin{figure}
\includegraphics[width=\columnwidth]{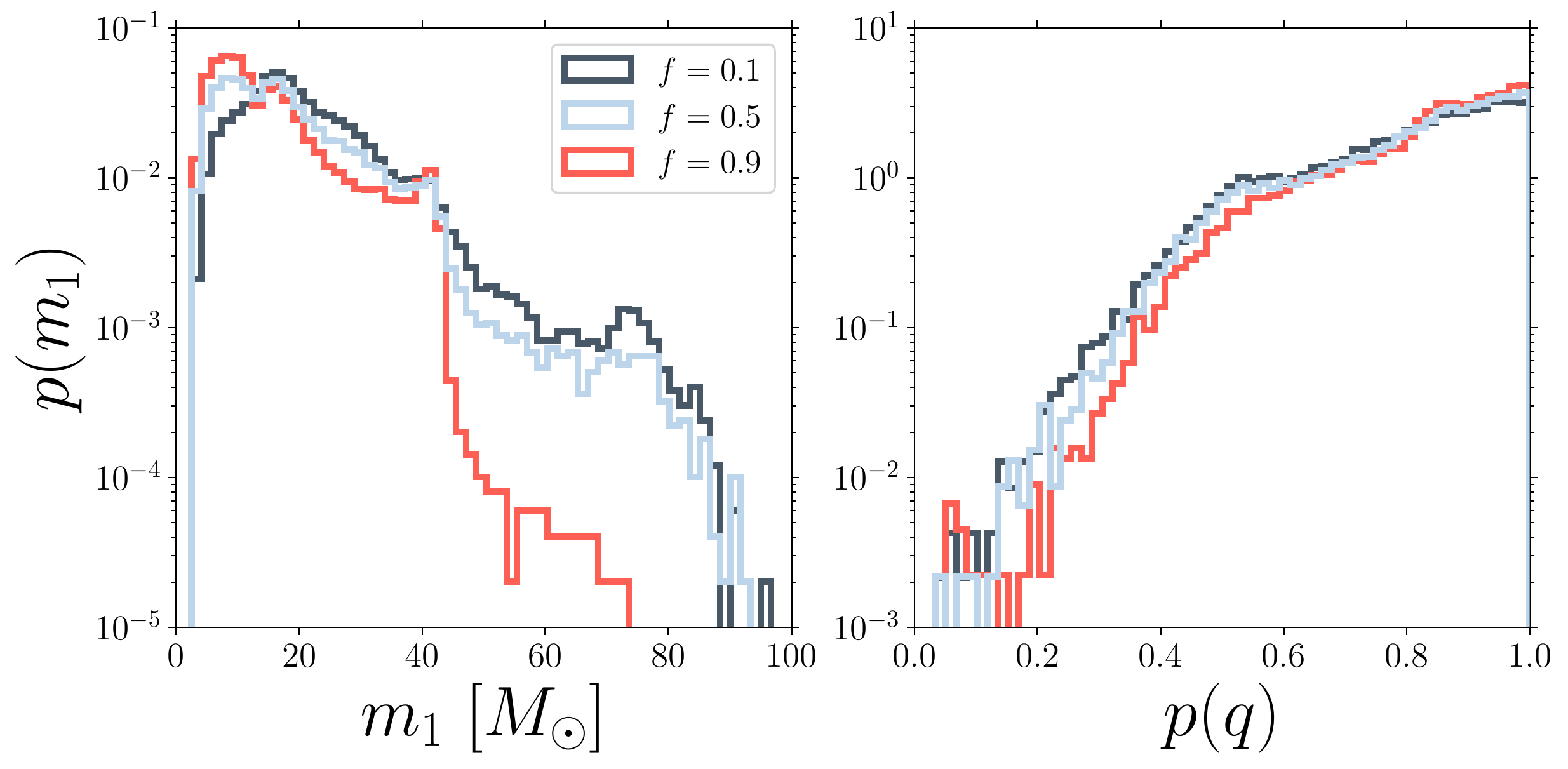}
\caption{The source-frame primary mass and mass ratio distribution for all merging BBHs in the mixture model.
We fix $\alpha=1$ and $r_v=1$ then mix the distributions with three values of the mixing fraction $f=[0.1,0.5,0.9]$
}
\label{fig:f_Distribution_plots}
\end{figure}

In the \texttt{CMC} models, the BH mass spectrum features three prominent peaks: the first at roughly $10-20\,M_{\odot}$ due to the assumptions concerning mass fallback during core collapse \citep[e.g.,][]{Fryer2012}, the second at roughly $40\,M_{\odot}$ due to the assumptions concerning the pair-instability \citep[e.g.,][]{Belczynski2016}, and the third at roughly $70-80\,M_{\odot}$ due to first-generation-BH-merger products retained in their host cluster post-merger \citep[e.g.,][]{Rodriguez:2019huv}. The first two peaks are also found in the \texttt{COSMIC} models 
because they are features of single star evolution assumptions, while the third peak is unique to the dynamical cluster environment.

The shift toward higher primary mass in the low $\alpha$ populations is due to an increased rate of stellar mergers during the common envelope phase, before a BBH forms, of lower mass BBH progenitors. Conversely, the shift towards lower primary masses with increasing common envelope ejection efficiency is a result of fewer stellar mergers during the common envelope phase. 
BH masses in \texttt{COSMIC} are directly correlated with progenitor core masses, thus lower mass BHs with have lower mass progenitors, which will enter common envelope evolution in tighter orbits. Since the delay times for merging BBHs in all of our models which undergo common envelope evolution are short enough for the majority of the population to merge within a few Gyr, models with higher $\alpha$ retain the low-mass BBH mergers while those with lower $\alpha$ retain relatively higher mass BBHs.


The shift toward higher primary masses at higher $r_v$ is a consequence of the effect of $r_v$ on BH cluster dynamics. 
Mass segregation arguments suggest that the most massive BHs in a cluster will, on average, be the first to be ejected from their host cluster and merge \citep[e.g.,][]{Morscher2015,Kremer2020}. Lower-mass ($\approx 10-15\,M_{\odot}$) BHs become dynamically active only after the most massive
BHs have been ejected. For smaller initial $r_v$, high-mass BHs are dynamically processed and
ejected early on. Therefore, in these clusters, high-mass ($M\gtrsim 40 M_{\odot}$) BHs, including the second-generation BHs with masses in the pair-instability gap, typically merge at high redshift leaving only the least massive BHs in any significant quantity at late times (low redshift). Meanwhile, for high-$r_v$ clusters, the initial relaxation time is longer \citep[e.g.,][]{BinneyTremaine1987}, so many high-mass BHs still remain and may merge at late times. Thus, BBH mergers tend to have higher component masses for higher $r_v$ clusters, as shown in Fig. \ref{fig:m1q_Distribution_plots}.

The most significant difference between the two formation channels is the existence or absence of BHs in the upper mass gap, with $m_1\gtrsim 45\,M_\odot$.
All stars in our simulations, including those formed in GCs, are subject to PISNe which results in a sharp mass cutoff near $m_1\approx 45\ M_{\odot}$ for all first generation BBH mergers.  We note that the assumptions for neutrino mass loss in our \texttt{COSMIC} and \texttt{CMC} models differ slightly such that the maximum mass for a first generation BBH component is $44.5\,\rm{M_{\odot}}$ in \texttt{COSMIC} and $40.5\,\rm{M_{\odot}}$ in \texttt{CMC}. This difference arises in the choice for neutrino mass loss to carry away a fixed $0.5\,\rm{M_{\odot}}$ or $10\%$ of the compact object mass at formation. Only hierarchical mergers, which do not occur in our isolated binary simulations, can result in BBHs with masses polluting the upper mass gap.
This fact serves as the primary means of disentangling formation channels and measuring their mixing fraction $f$.
Figure~\ref{fig:f_Distribution_plots}, for example, shows the \textit{total} $m_1$ and $q$ distributions resulting from combining field and cluster channels, assuming $\alpha=1$ and $r_{v}=1$ for several different mixing fractions $f$, the exact value of which sensitively controls the prevalence of high-mass systems.
Thus the relative numbers of high- and low-mass binaries among GWTC-2 serves as a sensitive probe of the 
value of $f$.

\begin{figure}[h]
\includegraphics[width=\columnwidth]{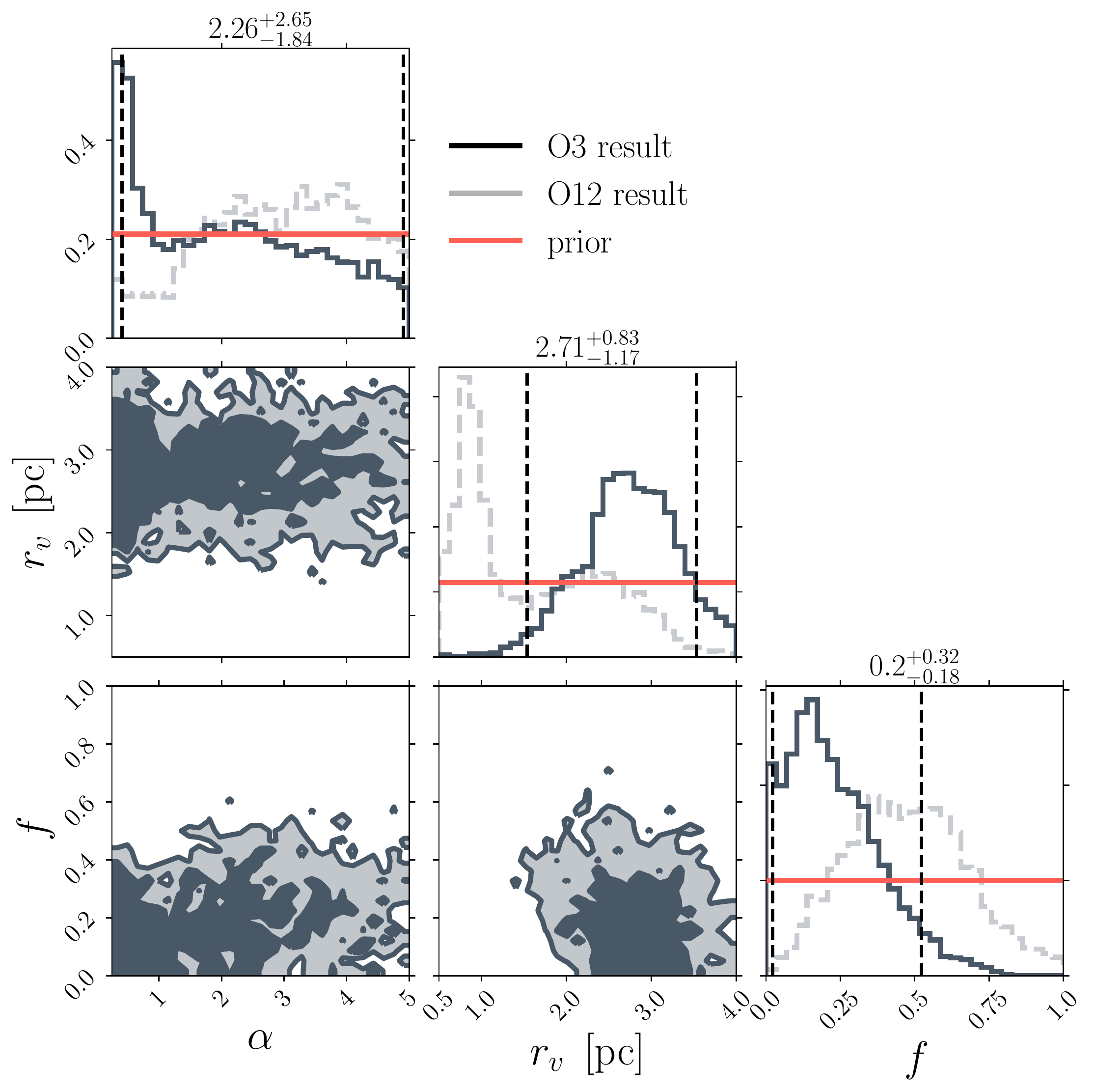}
\caption{The posterior distribution of our model inferred by using all BBHs up to GWTC-2.
The contours represent $50\%$ and $90\%$ credible bounds.
The grey lines in the 1D marginalized results are posterior distribution inferred by using only BBHs in GWTC-1.
}
\label{fig:result}
\end{figure}

\textbf{\textit{Inference results}}. 
We infer the hyper-parameters of Eq.~\eqref{eq:mixingFraction} using the posterior samples publicly released in support of GWTC-1~\cite{LIGOScientific:2018mvr,gwtc1_datarelease} and GWTC-2~\cite{Abbott:2020niy,gwtc2_datarelease}, restricting to the 44 events with false alarm rates $<1\,\mathrm{yr}^{-1}$.
Our main results are shown in Fig.~\ref{fig:result}.
As shown in Fig.~\ref{fig:m1q_Distribution_plots}, a component mass larger than 
$\sim45\ M_{\odot}$ is a signature of GC-formed BBHs. 
In GWTC-1, only GW170729 has a source-frame primary mass estimate with a median larger than $45\ M_{\odot}$.
With only 10 events in GWTC-1, the mixing fraction is therefore rather unconstrained, and the data show a preference for a near-equal mixing of the two populations instead of one dominating the other (grey $1-\rm{D}$ histograms).
This is not the case for GWTC-2. There are 8 events with median source-frame primary masses 
$>45\ M_{\odot}$.
Together with the events from GWTC-1, 9 out of 44 events pass the PISN mass gap.
This boost in number of high mass events significantly shifts the preferred value of the mixing fraction toward the cluster formation scenario.
Taking the median value of the posterior on the mixing fraction, $f=0.2$, we expect that $\sim 35$ of the 44 total GWTC-2 events arise from the cluster formation scenario.
Despite this preference for cluster formation, the field formation scenario can still contribute significantly to the entire observed population.
Our 90\% credible upper bound on the mixing fraction is $f=0.522$ suggesting that the contribution from the isolated channel could still surpass the cluster formation channel.

Our posterior on the common envelope efficiency $\alpha$ shows a mild preference towards lower values.
Our relative insensitivity to $\alpha$ is due to two effects.
First, varying $\alpha$ changes the mass distributions predicted by \texttt{COSMIC} by only a factor of a few.
Second, since the mixing fraction favours the cluster formation channel over the isolated formation channel,
the effective number of events which can be used to constrained the common envelope efficiency is small.

In contrast, the initial cluster virial radius $r_v$ is reasonably well-constrained, since 
our module suggests that more BBH mergers originate in GCs than in isolation.
There are 5 events with a component mass larger than $60\ M_{\odot}$, which make up more than $10\%$ of the entire dataset.
As clusters with larger initial virial radii tend to yield more massive mergers at higher redshift (including hierarchical mergers with masses in the pair-instability gap), the excess of high-mass events hints that the majority of clusters may have been born with moderately large virial radius (we predict a median value of $2.71\,$pc).

\begin{figure}[h]
\includegraphics[width=\columnwidth]{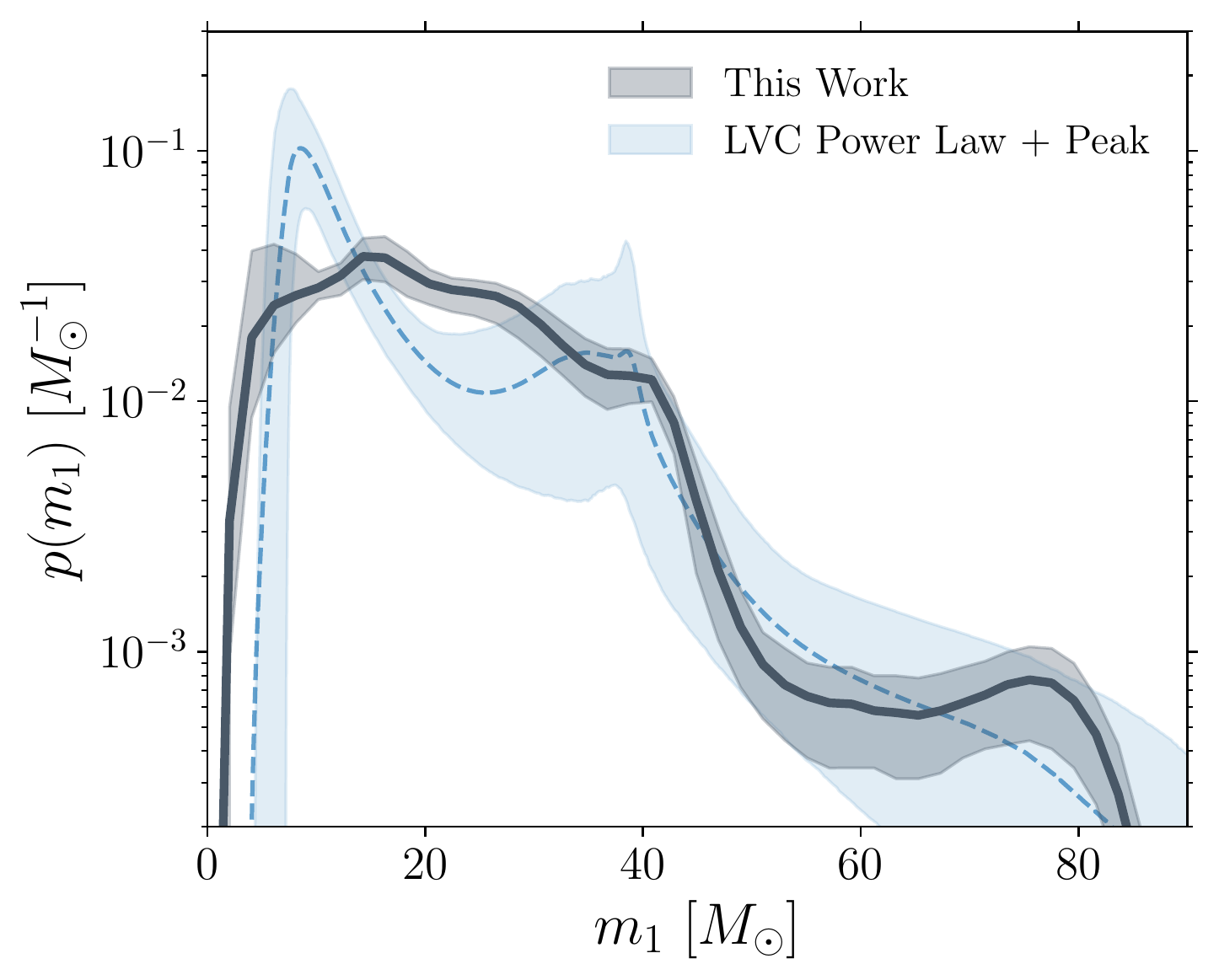}
\caption{The primary mass distribution (shaded grey) inferred by our mixture model.
Specifically, the shaded band shows the central 90\% credible bound on $p(m_1)$ as a function of primary mass, while the solid line marks the population predictive distribution: the inferred probability distribution on $m_1$ after marginalization over the field-cluster mixing fraction $f$, common envelope efficiency $\alpha$, and initial cluster virial radius $r_v$.
For comparison, the shaded blue band marks the analogous result obtained by the LVC under their ``Power Law + Peak'' mass model; the dashed blue line is the corresponding population predictive distribution.
}
\label{fig:result_realization}
\end{figure}

Figure~\ref{fig:result_realization} shows our inferred primary mass distribution, marginalized over our posterior on $f$, $\alpha$, and $r_v$.
For comparison, the light blue band indicates the mass spectrum inferred by the LVC using their phenomenological ``Power Law + Peak'' model, 
where the primary mass distribution is modeled as the superposition of a power law and a Gaussian~\cite{Abbott:2020gyp}.
Despite their radically different parametrizations (responsible for the systematic disagreement at low masses), both models yield qualitatively similar structure at moderate and high masses: a steep drop near $\sim 45\,M_\odot$ (corresponding to the onset of the pair instability mass gap in our model) and a plateau extending to $80\,M_\odot$.
Our results suggest that, in the future, phenemenological mass models should be designed that can more readily capture features like these; despite their flexibility, the current models employed in Ref.~\cite{Abbott:2020gyp} cannot natively accommodate both sharp cutoffs and shallow high-mass plateaus.


\textbf{\textit{Discussion}}. Previous analyses \citep[e.g.,][]{Kremer2019,Kremer2020} have shown a cluster's initial $r_v$ plays a prominent role in the cluster's long-term dynamical evolution. In particular, GCs born with smaller initial $r_v$ (and thus shorter relaxation times), are more likely to have undergone core collapse by the present day. In this case, the relative ratio of core-collapsed to non-core-collapsed GCs observed at present may provide a complementary constraint upon the initial $r_v$ distribution \citep{Kremer2020}. Taking the Milky Way GCs as a representative sample, roughly $80\,\%$ of clusters have well-resolved cores (i.e., are non-core-collapsed) at present \citep{Harris1996}, potentially hinting that relatively large initial $r_v$ are typical, consistent with the predictions from our GWTC-2 inference.

There are a number of assumptions made in this work that can be improved upon in the future.
To safeguard against systematic model uncertainties, we do not include merger rates in our inference.
For completeness, 
we report the comoving merger rate at $z < 0.1$ under each formation channel and hyper-parameter choice in Table~\ref{Tb:rates}. The local merger rates from the isolated formation channel are 
lie near the maximum value of the $90\%$ confidence intervals of the reported 
merger rates inferred from GWTC-2 of $19.7^{+57.3}_{-15.9}\,\rm{Gpc}^{-3}\,\rm{yr}^{-1}$. This illustrates a potential systematic bias in the other parameters which define the isolated binary formation models. In particular, \texttt{COSMIC} assumes that natal kicks for BHs are drawn from a Maxwellian distribution with $\sigma=265\,\rm{km/s}$ and then weighted by the amount of mass that falls back onto the proto-compact-object during formation \cite{Fryer2012}. If stronger kicks are assumed, BBHs with lower masses will preferentially be unbound, altering both the merger rate and mass distributions in the isolated formation scenario \citep[e.g.][]{Belczynski2002, Wong:2019uni,Belczynski2020}. 
\texttt{COSMIC} also assumes that stable Roche-overflow mass transfer is conservative, which may over 
predict the orbital evolution of a mass transferring binary \citep[e.g.][]{vanSon2020}.
This is further compounded by our uncertain choice, following \cite{Belczynski2008}, of the mass ratios for which mass transfer is assumed to be dynamically unstable leading to a common envelope.
If mass transfer proceeds stably for a wider range of mass ratios, fewer systems will experience dramatic orbital tightening during common envelope, potentially lowering predicted merger rates.
We leave a full study of the combined effects of BH natal kicks, common envelope, mass transfer stability, and accretion efficiency to a future study. 

\begin{table}
\caption{Event rates for each hyper-parameter and formation scenario. Rates are given for $z < 0.1$ and in comoving volume per time.}
\begin{tabularx}{\columnwidth}{ccccccccc}
\hline
\hline
Isolated binaries &&&&&&&&\\
\hline

\hline
$\alpha$ & 0.25 & 0.5 & 0.75 & 1 & 2 & 3 & 4 & 5 \\ 
$\Gamma_{\alpha}$\,[Gpc$^{-3}$yr$^{-1}$] & 63.8 & 62.5 & 66.5 & 72.2 & 76.6 & 75.1 & 73.9 & 72.1\\

\hline
Globular clusters &&&&&&&&\\
\hline
\hline
$r_v\,[\rm{pc}]$ & \multicolumn{2}{c}{0.5} & \multicolumn{2}{c}{1} & \multicolumn{2}{c}{2} & \multicolumn{2}{c}{4} \\ 
$\Gamma_{r_v}$\,[Gpc$^{-3}$yr$^{-1}$] & \multicolumn{2}{c}{31.6} & \multicolumn{2}{c}{26.8} & \multicolumn{2}{c}{20.9} & \multicolumn{2}{c}{8.7} \\

\hline
\hline
\end{tabularx}
\label{Tb:rates}
\end{table}

Although the rate estimates from clusters shown in Table \ref{Tb:rates} match well the reported comoving rate inferred from GWTC-2, the cluster estimates also feature several uncertainties. A major uncertainty is the assumed cluster formation history. Observations of young stellar clusters in the local universe \citep[e.g.,][]{PortegiesZwart2010} indicate that initial cluster virial radii are roughly independent of cluster birth time and metallicity, however this is highly uncertain. If $r_v$ does vary with cluster birth time, it may substantially affect the conclusions of this study. Furthermore, the cluster birth time distribution itself is highly uncertain. Current theories of cluster formation fall into two main categories: clusters formed through active star formation \citep[e.g.,][]{ Kruijssen2015,ElBadry2019} and clusters formed due to the collapse of dark matter halos during or before the epoch of reionization \citep[e.g.,][]{FallRees1985}. Here, we have assumed the former scenario but if a large population of present-day GCs were instead born during reionization, our results may again change significantly. Future work should consider more carefully the various possible assumptions regarding cluster formation scenarios and the dependence of cluster properties such as $r_v$ on these various scenarios. 

Since spin is not self consistently calculated in our simulations, we only use binary component masses and merger redshifts in this study.
Given that we are beginning to robustly resolve features in the BBH spin distribution~\cite{Abbott:2020gyp},
including spin information is an obvious improvement that we plan to carry out in future studies.
In the future, constraints on orbital eccentricities may additionally offer a powerful means of discriminating between field and dynamical formation scenarios~\cite{Rodriguez2018,Samsing_2018,Romero_Shaw_2019}.
Lastly, due to the still moderate number of events, we chose to exclude other proposed formation channels such as primordial BHs, nuclear star clusters, open clusters, stellar triples, and AGN disks. 
This simplification has strong implications, especially when considering the relative contribution of each channel \cite{Zevin2020inprep}.
As future observational runs provide more events, and theoretical predictions for formation channels mature, we hope be able to jointly constrain more formation channels at once.

To conclude, we have extracted the mixing fraction of BBHs formed in isolated binaries and GCs while simultaneously placing constraints on the channel-specific hyper-parameters,
notably the initial virial radius of GCs where the BBHs were formed.
Our work marks an important milestone of learning astrophysics from populations of observed GW events.
Instead of constraining a physics-agnostic phenomenological model,
we use a population of GW events to place constraints on physical parameters such as mixing fraction directly.
Furthermore, our model suggests that the GC properties inferred from the detected GW population are consistent with electromagnetic observations of present-day GCs, notably cluster core radii distributions.
This result may be readily compared with other independent measurements to test theories of GC formation and evolution.
As the number of GW detections increases in the future and theoretical models improve, one can apply the same methodology to discover and test a plethora of astrophysical theories more precisely.

\textbf{\textit{Acknowledgement}}. The authors thanks Vishal Baibhav, Emanuele Berti, Will Farr, Vicky Kalogera, Ken K. Y. Ng, Isobel Romero-Shaw, and Mike Zevin for constructive feedback.
K.W.K.W. is supported by NSF Grants No. PHY-1912550 and AST-2006538, NASA ATP Grants No. 17-ATP17-0225 and 19-ATP19-0051, and NSF-XSEDE Grant No. PHY-090003. 
This research has made use of data, software and/or web tools obtained from the Gravitational Wave Open Science Center (https://www.gw- openscience.org), a service of LIGO Laboratory, the LIGO Scientific Collaboration and the Virgo Collaboration.
This research project was conducted using computational resources at the Maryland Advanced Research Computing Center (MARCC). The authors would like to acknowledge networking support by the GWverse COST Action CA16104, ``Black holes, gravitational waves and fundamental physics.
KK is supported by an NSF Astronomy and Astrophysics Postdoctoral Fellowship under award AST-2001751.
The Flatiron Institute is supported by the Simons Foundation.
Portions of this work were performed during the CCA LISA Sprint, supported by the Simons Foundation.

\appendix*
\setcounter{equation}{0}
\section{Appendix: hierarchical Bayesian modelling with normalizing flow}

In this section, we give a brief summary of the hierarchical inference and deep learning methods used in this work.
We refer the readers to the comprehensive descriptions in e.g. Refs~\cite{Wong:2019uni,Wong:2020jdt,Mandel:2018mve} for additional details.

Given data $\bm{d}$ spanning a number $N_{\rm obs}$ of gravitational-wave detections, we wish to infer the posterior $p(\bm{\lambda}|\bm{d})$ on the hyperparameters $\bm{\lambda}$ governing their population.
Assuming that the population is described as an inhomogenous Poisson process, the posterior takes the form~\cite{Loredo2004,Mandel:2018mve}

\begin{align}
p({\bm \lambda}|{\bm d}) &\propto \ \pi({\bm \lambda})  \,e^{- \alpha({\bm\lambda}) N({\bm \lambda})} N({\bm \lambda})^{N_{\rm obs}} \nonumber \\ &\times\prod_{i=1}^{N_{\rm obs}} \! \int \!
    \frac{ p({\bm d}|{\bm \theta_i})\, p_{\rm pop}({\bm \theta_i}|{\bm \lambda}) }{ \alpha({\bm \theta_i}) } {\rm d}{\bm \theta_i} \,,
\label{eq:posterior}
\end{align}

where $N(\bm{\lambda})$ is the intrinsic volume-integrated event rate predicted by the model, and $N_{\rm obs}$ is the number of observed event in the data.
$p({\bm d}|{\bm \theta_i})$ is the likelihood of the $i$-th event in the observed catalogue and $\pi({\bm \lambda})$ is the prior on our hyperparameters.
Meanwhile, $p_{\rm pop}({\bm \theta_i}|{\bm \lambda})$ is the probability density function for the event-level parameters ${\bm \theta_i}$; this function will be computed using a normalizing flow emulator.

The term $\alpha(\bm \lambda)$ is known as the selection bias, and gives the fraction of events one expects to detect:
\begin{align}
\alpha(\bm \lambda) = \int p_{\rm pop}({\bm \theta'}|{\bm \lambda}) P_{\rm det}(\bm \theta') d {\bm \theta'}.
\label{eq:selectionFunction}
\end{align}
where $P_{\rm det}(\bm{\theta'})$ is the probability that an event with specific parameters ${\bm \theta'}$ is successfully detected.
In principle, one needs to inject a large amount of signals and recover them with a search pipeline to estimate the selection bias, which is very computationally expensive.
In practice, we follow the procedure described in \cite{Abbott:2020gyp,2019RNAAS...3...66F}, reweighting an injection campaign done by the LVKC to compute the selection bias for the ${\rm O12}+{\rm O3a}$ catalog.
We interpolate $N({\bm \lambda})$ and $\alpha({\bm \lambda})$ and use the interpolated function during the inference to maximize computational efficiency.

As discussed above, the absolute merger rates predicted by the \texttt{COSMIC} and \texttt{CMC} simulations are likely subject to unknown systematic uncertainties.
We therefore marginalize the posterior shown in Eq.\eqref{eq:posterior} over the intrinsic merger rate, using a prior $\pi(N) \propto 1/N$.
Additionally, we do not know the underlying likelihood $p({\bm d}|{\bm \theta_i})$ for each catalog event, but only the \textit{posterior} $p({\bm \theta_i}|{\bm d})$ obtained under some default prior $\pi({\bm \theta_i})$ adopted by the LVKC during parameter estimation.
Under both of these conditions, Eq.~\eqref{eq:posterior} can be written as~\cite{2018ApJ...863L..41F}
\begin{equation}
p({\bm \lambda}|{\bm d}) \!\propto\! \pi({\bm \lambda})  \prod_{i=1}^{N_{\rm obs}} \! \int \!\frac{p({\bm \theta_i}|{\bm d})}{\pi({\bm \theta_i})}
\frac{p_{\rm pop}({\bm \theta_i}|{\bm \lambda})}{\alpha(\bm \lambda)}
{\rm d}{\bm \theta_i} \,.
\label{eq:posteriormarg}
\end{equation}
Specifically, the LVKC releases their event posterior in the form of discrete samples produced by their parameter estimation pipeline.
Given these samples, the integral in Eq.~\eqref{eq:posteriormarg} can be evaluated by using importance sampling, which turns the integral into a discrete sum over the event posterior PDF samples
\begin{align}
p(\bm{\lambda}|\bm{d}) \propto  \pi(\bm{\lambda})\prod_{i=1}^{N_{\rm obs}}\frac{1}{S_i}\sum_{j=1}^{S_i} \frac{p_{\rm pop}(^j\bm{\theta}_i|\bm{\lambda})}{\pi(^j\bm{\theta}_i)\alpha({\bm \lambda})}.
\label{eq:populationPosterior_discrete}
\end{align}
where $j$ labels the posterior samples of the $i$-th event and $S_i$ is the total number of samples per event.
The default prior used in the LVKC \texttt{LALInference} software~\cite{2015PhRvD..91d2003V} is uniform in \textit{detector-frame} masses and quadratic in luminosity distance, whereas we wish to model the distributions of source-frame masses and redshifts.
In terms of these desired coordinates, the \texttt{LALInference} prior takes the form
\begin{align}
\pi(m_1,m_2,z) \propto (1+z)^2 D_L^2(z) \left( D_c(z) + \frac{c(1+z)}{H(z)}\right).
\end{align}
To evaluate Eq.~\eqref{eq:populationPosterior_discrete}, we make use of the posterior samples presented in Refs.~\cite{LIGOScientific:2018mvr} and \cite{Abbott:2020niy} and released through the Gravitational-Wave Open Science Center~\cite{gwosc}.
In particular, we use the ``\texttt{PublicationSamples}'' dataset associated with each event, and restrict to the 44 events with false alarm rates $<1\,\mathrm{yr}^{-1}$ following Ref.~\cite{Abbott:2020gyp}.

The final term we need for evaluating Eq.~\eqref{eq:populationPosterior_discrete} is the distribution of observables predicted by our simulation $p_{\rm pop}({\bm \theta}|{\bm \lambda})$,
which we obtain through training a deep learning emulator on the simulations.
Here we give the network-related hyperparameters for reproducing our result.
A detailed discussion of the architecture of the neural network and of the training procedure can be found in Refs.~\cite{Wong:2019uni,Wong:2020jdt}.
We train a masked autoregressive flow network~\cite{2017arXiv170507057P} with 10 hidden layers, each layer having 1024 units with ReLU activation. 
We include 3 observables $\{m_1,m_2, z\}$ and 3 hyper-parameters $\{f,\alpha,r_v\}$ in our training,
as tabulated in Table~\ref{Tb:parameters}.
The training set contains 160 simulations with different combinations of 8 values of $\alpha \in [0.25,5]$, 4 values of $r_v \in [0.5,4]$ and 5 values of $f \in [0,1]$.
Each simulation has $10^5$ events, of which $80\%$ are randomly chosen our training set and $10\%$ for both validation and test sets.
Note that we follow the LVKC convention to enforce $m_1 > m_2$.
We train the network for 100 epoch on a Nvidia K80 GPU to ensure convergence.
The code for the neural network is written in python with {\sc pytorch}~\cite{paszke2017automatic}.
Eq.~\eqref{eq:populationPosterior_discrete} is then sampled using the MCMC package \texttt{emcee}~\cite{2013PASP..125..306F} to produce the results shown in this work.

\bibliography{main}

\end{document}